\documentclass[prd,preprint,showpacs]{revtex4}

\usepackage{amssymb,amsmath,graphicx}

\newcommand{\II}{{\boldsymbol{1}}}
\newcommand{\RR}{{\mathbb R}}
\newcommand{\GG}{{\mathcal G}}
\newcommand{\xb}{{\boldsymbol{x}}}
\newcommand{\ip}[2]{{\langle #1\mid #2\rangle}}

\begin{document}

\title{Probability distributions of smeared quantum stress tensors}

\begin{flushright}
ESI 2250
\end{flushright}

\author{Christopher J. Fewster}
\email{cjf3@york.ac.uk}
\affiliation{Department of Mathematics, University of York, 
Heslington, York YO10 5DD,
United Kingdom}

\author{L. H. Ford}
\email{ford@cosmos.phy.tufts.edu}
\affiliation{Institute of Cosmology, Department of Physics and
  Astronomy, 
Tufts University, Medford, Massachusetts 02155, USA}
\author{Thomas A. Roman}
\email{roman@ccsu.edu}
\affiliation{
Department of Mathematical Sciences, Central Connecticut State
University, 
New Britain, Connecticut 06050, USA}

\date{\today}

\begin{abstract}
We obtain in closed form 
the probability distribution for individual measurements of
the stress-energy tensor of two-dimensional conformal field theory
in the vacuum state, smeared in time
against a Gaussian test function.  The result 
is a shifted Gamma distribution
with the shift given by the previously known optimal
quantum inequality bound. For small values of the central charge it 
is overwhelmingly likely that individual measurements of the
sampled energy density in the vacuum give negative results. For the
case of a single massless scalar field, the probability of finding a
negative value is $84\%$.
We also report on computations for
four-dimensional massless scalar fields showing that the
probability distribution of the smeared 
square field is also a shifted Gamma distribution, but that the 
distribution of the energy density is not. 
\end{abstract}

\maketitle

\baselineskip=14pt

The expectation value of the stress tensor of a quantum field, when 
suitably renormalized, can be used to give a description of the
gravitational effects of quantum fields. However, this semiclassical
theory clearly has limitations in that it does not describe
fluctuations around the mean value. In recent years, several authors
have discussed possible physical effects of these fluctuations.
For recent reviews with further references, see 
Refs.~\cite{Stochastic,FW07}.
Much of the work on this topic has
involved seeking observables which can be related to integrals of the
stress tensor correlation function. However, this amounts to
calculating a variance, or a second moment, of the probability
distribution for measurement results of the smeared stress-energy tensor. 
It is desirable to go further and determine the complete
distribution function. In this paper, we will report results for such
a function.

Here we will be concerned with Minkowski spacetime in the vacuum
state and seek the probability distribution function for a stress
tensor operator which has been averaged with a sampling function. This
averaging is crucial to having a well-defined probability
distribution. (A recent attempt has been made to assign a distribution
to a local, unsmeared, stress tensor operator~\cite{Duplanic}. 
However, in our view
this is not a meaningful quantity.) There is a deep connection between
the probability distribution for vacuum fluctuations, and the quantum
inequalities (QIs) which place lower bounds on the expectation values of
smeared stress tensors in arbitrary physically reasonable quantum 
states~\cite{F78,F91,FR95,FR97,Flanagan97,FewsterEveson98,Fe&Ho05}. 
These lower bounds
represent the lowest eigenvalues of the smeared operators. 
As a consequence of the Reeh--Schlieder theorem (see, e.g.,~\cite{Haag}), they are also 
the lower bounds of the support of the corresponding probability 
distributions in the vacuum state~\footnote{The connection begins
with the observation that the following three statements are equivalent,
where $A$ is a self-adjoint operator (with domain $D(A)$ and spectrum $\sigma(A)$) 
representing a quantum observable:
(a) $\ip{\psi}{A\psi} \ge -Q$ for all normalised $\psi\in D(A)$; (b)
$\sigma(A) \subset [-Q,\infty)$; (c) the probability measure $\mu_\psi$ that
describes the distribution of individual measurement results of $A$ in state $\psi$ has
support  $\textrm{supp}\,\mu_\psi \subset [-Q,\infty)$. [The implication (a)$\implies$(b)
is the Rayleigh--Ritz variational principle;  (b)$\implies$(c) 
because measurements always give values in the spectrum; (c)$\implies$(a) because 
any random variable taking values in $[-Q,\infty)$ has its expected value in that range.]
%
In general, $\textrm{supp}\,\mu_\psi$ (for
a fixed choice of $\psi$) can be a proper subset of $\sigma(A)$, and in particular
the infimum of the support of $\mu_\psi$ can exceed $\inf \sigma(A)$.  For example,
this happens if $\inf\sigma(A)$ is an isolated eigenvalue whose eigenfunction is orthogonal to $\psi$.
However, the spectrum and the support of the probability measure coincide if $A$ is a local
observable, such as a smeared field, and $\psi$ is the vacuum state $\Omega$ of a Minkowski space QFT 
obeying the standard assumptions of algebraic quantum field theory \cite{Haag}. To see this, note first that
if $\lambda\in\sigma(A)$ then $P_\epsilon$,
the spectral projection of $A$ corresponding to the interval $(\lambda-\epsilon,\lambda+\epsilon)$, is nonzero
for any $\epsilon>0$. As $P_\epsilon$ is a local observable (because $A$ is) the Reeh--Schlieder
theorem entails that it cannot annihilate the vacuum, and hence there is a nonzero probability
$\|P_\epsilon\Omega\|^2$ of obtaining a measurement of $A$ in the interval 
$(\lambda-\epsilon,\lambda+\epsilon)$, which must therefore intersect the support of $\mu_\Omega$.
Taking $\epsilon\to 0$ we conclude that $\lambda\in\textrm{supp}\,\mu_\Omega$ (which is a closed subset of $\RR$).
Varying $\lambda$, this argument establishes that $\sigma(A)\subset
\textrm{supp}\,\mu_\Omega$, and we already know that the reverse inclusion holds [by the implication
(b)$\implies$(c) above].
In particular, $\sigma(A)$ and $\textrm{supp}\mu_\Omega$ have a common infimum, which provides a sharp quantum
inequality bound $\ip{\psi}{A\psi}\ge \inf\textrm{supp}\,\mu_\Omega$ for all normalized $\psi\in D(A)$ by our earlier observations.}. 
In this paper, we will
concentrate on conformal field theory (CFT) in two spacetime dimensions, but 
will also report a result for the square of a massless scalar field in four
spacetime dimensions. Further details will be given in a later
paper.

We consider a general unitary, positive energy CFT with central charge $c$ in two-dimensional
Minkowski space (with $+-$ signature and inertial coordinates $x^a$). A general reference 
is~\cite{FurlanSotkovTodorov89}. Our notation is as follows: $T_L$ and
$T_R$ are the left- and right-moving components of the (renormalized)
stress-energy tensor, so that 
\begin{eqnarray}
T_{00}(x^0,x^1) = -T_{11}(x^0,x^1) &=& T_R(u) + T_L(v)  \nonumber \\
T_{01}(x^0,x^1) = T_{10}(x^0,x^1) &=& -T_R(u) + T_L(v)
\end{eqnarray}
where $u=x^0-x^1$, $v=x^0+x^1$. As $T_R$ and $T_L$ commute, we can find
the probability distribution for a smearing of $T_{00}$ from the probability
distributions for corresponding smearings of $T_L$ and $T_R$. We
begin by considering $T_R$ only and dropping the subscript 
(identical results hold for $T_L$). The vacuum $n$-point functions will 
be denoted $G_n(u_n,\ldots,u_1) = \langle T(u_n)\cdots T(u_1)\rangle$, 
with the convention that  $G_0 = \langle \II\rangle = 1$; where 
one or more of the $u_j$ is replaced by the letter 
$f$, this indicates smearing against a test function $f(u_j)$,
and we write $\GG_n[f]=
G_n(f,\ldots,f) = \langle T(f)^n\rangle$ for the $n$-th moment of $T(f)$
in the vacuum state. 
All test functions will be assumed real-valued and rapidly decaying at infinity.

Our goal is to determine an expression for the moment generating function
\begin{equation}
M[\mu f] = \sum_{n=0}^\infty \frac{\mu^n\GG_n[f]}{n!}
\end{equation}
from which the underlying probability distribution can be determined under
suitable conditions. Here, we take $\int f(u)\,du$ to be dimensionless and
$\mu$ with dimensions of $\text{\em length}^{2}$. We start with the 
Ward identities in CFT (see, e.g., Sect.~3 of
\cite{FurlanSotkovTodorov89}), which entail the recurrence relation
\begin{eqnarray}\label{eq:Ward}
&{}& \!\!\! G_n(u_n,\ldots,u_1) = \sum_{j=1}^{n-1}\left[
\frac{c}{8\pi^2}\frac{G_{n-2}(u_{n-1},\ldots,
\hat{u}_j,\ldots,u_1)}{(u_n-u_j-i0)^4}\right. \nonumber\\
 &-&\frac{\partial_j G_{n-1}(u_{n-1},\ldots,
u_1)}{2\pi(u_n-u_j-i0)} \left.
-\frac{G_{n-1}(u_{n-1},\ldots,u_1)}{\pi(u_n-u_j-i0)^2}\right],
\end{eqnarray}
where the hat denotes an omitted variable. Since $G_0=1$ and
$G_1(u_1)=\langle T(u_1)\rangle \equiv 0$, it follows immediately that
\begin{equation}
G_2(u_2,u_1) = \frac{c}{8\pi^2(u_2-u_1-i0)^4}.
\end{equation}
Smearing Eq.~\eqref{eq:Ward} against $n$ copies of $f$
and noting that the first line
of the right-hand side provides $n-1$ identical
contributions, we have
\begin{equation}\label{eq:Ward2}
\GG_n[f] = (n-1) \GG_2[f]\GG_{n-2}[f] + \sum_{j=1}^{n-1} I_{n,j} \,,
\end{equation}
where it may be shown that
\begin{equation}
I_{n,j} = G_{n-1}(f,\ldots,\underbrace{f\star f}_j,\ldots,f),
\end{equation}
with the underbrace indicating that $f\star f$ is
inserted in the $j$'th position from the right, and $f\star f$ given 
after some computation by
\begin{equation}
f\star f(u)= \int_{-\infty}^\infty dw\, \frac{f(w)f'(u)-f'(w)f(u)}{2\pi(w-u)}.
\end{equation}
Note that the integrand is nonsingular at $w=u$. In deriving this
formula several steps of integration by parts have been used; no 
boundary terms are introduced provided $f$ decays sufficiently fast at
infinity.

Now suppose that a one-parameter family $f_\lambda(u)$
($\lambda\in\RR)$ 
can be found, satisfying the differential equation
\begin{equation}
\frac{df_\lambda}{d\lambda} = f_\lambda\star f_\lambda
\label{eq:diff}
\end{equation}
and initial condition $f_0=f$ ($\lambda$ has dimensions of $\text{\em length}^2$). Then, replacing $f$ by $f_\lambda$ in 
Eq.~\eqref{eq:Ward2}, the sum over $j$ can be rewritten as a derivative
with respect to $\lambda$ \footnote{For example, 
use $f_{\lambda+\delta} = f_\lambda + \delta f_\lambda\star f_\lambda +
O(\delta^2)$ and linearity of $G_{n-1}$ in each slot.}, yielding 
\begin{equation}
\GG_n[f_\lambda] = (n-1)\GG_2[f_\lambda]\GG_{n-2}[f_\lambda]+
\frac{d}{d\lambda}\GG_{n-1}[f_\lambda].
\end{equation}
Multiplying by $\mu^{n-1}/(n-1)!$ and summing over $n\ge 2$,
this becomes a partial differential equation 
\begin{equation} \label{eq:M_diff_eqn}
\frac{\partial}{\partial \mu} M[\mu f_\lambda]
= \mu\GG_2[f_\lambda] M[\mu f_\lambda] + 
\frac{\partial}{\partial\lambda} M[\mu f_\lambda] 
\end{equation}
for $M[\mu f_\lambda]$, on using the facts
$\GG_0[f_\lambda]=1$, $\GG_1[f_\lambda]=0$, and 
$\GG_n[\mu f_\lambda]=\mu^n \GG_n[f_\lambda]$. To solve this,
we write $M[f]=\exp (W[f])$, whereupon Eq.~\eqref{eq:M_diff_eqn} becomes 
\begin{equation}
\frac{\partial}{\partial \mu} W[\mu f_\lambda] = 
\mu\GG_2[f_\lambda]+\frac{\partial}{\partial\lambda} W[\mu f_\lambda]\,.
\label{eq:Weq}
\end{equation}
As $W[\mu f_\lambda]=0$
for $\mu=0$,  the unique solution is
\begin{equation}
W[\mu f_\lambda] = \int_{\lambda}^{\lambda+\mu} d\lambda'\, 
(\mu+\lambda-\lambda')\GG_{2}[f_{\lambda'}] \,,
\end{equation}
and we set $\lambda=0$ to obtain 
\begin{equation}\label{eq:W_final}
W[\mu f] = \int_0^\mu d\lambda\, (\mu-\lambda)\GG_{2}[f_{\lambda}].
\end{equation}
Equation~\eqref{eq:W_final} is similar to a formula derived in the 
Euclidean holomorphic picture
(in contrast to our Lorentzian treatment) 
by Haba in Eq.~(9) of~\cite{Haba90}. 

The differential equation Eq.~\eqref{eq:diff} can be solved
explicitly in the case of a
Gaussian sampling function $f(u) = e^{-u^2/\tau^2}/(
\tau\sqrt{\pi})$. Making the {\em ansatz} 
$f_\lambda(u) = A(\lambda)f(u)$, the calculation
\begin{equation}
(f_\lambda\star f_\lambda)(u) =
\frac{A(\lambda)^2}{\tau^3\pi^{3/2}}e^{-u^2/\tau^2} =
\frac{A(\lambda)}{\tau^2\pi} f_\lambda(u),
\end{equation}
reduces Eq.~\eqref{eq:diff} to 
$dA/d\lambda =  A(\lambda)^2/(\tau^2\pi)$, whose 
unique solution with  $A(0)=1$ is
\begin{equation}
A(\lambda) = \frac{\pi\tau^2}{\pi\tau^2-\lambda}.
\end{equation}
We now have $\GG_2[f_\lambda]=A(\lambda)^2\GG_2[f]$, so
\begin{align}
W[\mu f] &= \GG_{2}[f] \int_0^\mu d\lambda\, (\mu-\lambda)
A(\lambda)^2 \nonumber \\
&= \GG_{2}[f] \left[\tau^4\pi^2
\log\left(\frac{\pi\tau^2}{\pi\tau^2-\mu}\right)-
\mu\tau^2\pi\right] \nonumber \\ 
& = \frac{c}{24}\log\left(\frac{\pi\tau^2}{\pi\tau^2-\mu}\right)
- \frac{c\mu}{24 \pi\tau^2}\,,
\end{align}
where we have used
\begin{equation}
\frac{1}{(v-i\epsilon)^4} = \frac{1}{6}\int_0^\infty d\omega \,\omega^3
e^{-i\omega(v-i\epsilon)}
\end{equation}
to compute
\begin{equation}
\GG_{2}[f]  = \frac{c}{48\pi^2}\int_0^\infty d\omega
\,\omega^3 e^{-\omega^2\tau^2/2} = \frac{c}{24\pi^2\tau^4}.
\end{equation}
The moment generating function for the Gaussian-averaged
right-moving flux is then
\begin{equation}
M[\mu f] = e^{W[\mu f]} =
\left[\frac{e^{-\mu/(\pi\tau^2)}}
{1-\mu/(\pi\tau^2)}\right]^{c/24}.
\end{equation}
It remains to determine the corresponding probability distribution
$P(\omega)$ such that
\begin{equation}
M[\mu f] = \int_{-\infty}^\infty d\omega\, P(\omega) e^{\mu\omega}.
\end{equation}
As the moments obey the uniqueness conditions in the Hamburger moment
theorem~\cite{Reed--Simon}, there can be only one such distribution.  
(One needs $|a_n|\le CD^n n!$ for the
$n$'th moment $a_n=d^n M[\mu f]/d\mu^n|_{\mu=0}$, which is satisfied for our case.) 
It is a shifted Gamma distribution, of form
\begin{equation}
P(\omega) =
\vartheta(\omega+\omega_0) 
\frac{\beta^{\alpha}(\omega+\omega_0)^{\alpha-1}}{\Gamma(\alpha)} 
\exp(-\beta(\omega+\omega_0)),
\label{eq:shifted_Gamma}
\end{equation}
with parameters 
\begin{equation}
\omega_0 = \frac{c}{24\pi\tau^2},\qquad
\alpha = \frac{c}{24}, \qquad
\beta = \pi\tau^2,
\end{equation}
and where $\vartheta$ is the Heaviside 
function.  Note that Eq.~(\ref{eq:shifted_Gamma}) is singular at its
lower limit provided that $c < 24$, but the singularity is integrable,
so that the integrated probability is unity, as required. 


As described above, the lower bound 
of this distribution is also the best possible quantum inequality bound
on the expectation value in a normalized state $\psi$, so
\begin{equation}
\langle T(f)\rangle_\psi \ge -\omega_0
\end{equation}
for all physically reasonable $\psi$.
 This can be checked against
the optimal bound given in Theorem~4.1 of~\cite{Fe&Ho05}, 
\begin{equation}
\langle T(f)\rangle_\psi\ge  
-\frac{c}{12\pi}\int \left(\frac{d}{du} \sqrt{f(u)}\right)^2\,du =
-\frac{c}{24\pi\tau^2} = -\omega_0.
\end{equation}
Thus we have the expected agreement between the two methods. The
nonzero negative lower bound should be a general feature of stress
tensor probability distributions, and is required by the vanishing
mean value in the vacuum state, in contrast to the result claimed 
in Ref.~\cite{Duplanic}.

The cumulative distribution function of our probability distribution is 
an incomplete Gamma function. Perhaps surprisingly, 
it is overwhelmingly likely that an individual measurement will yield a negative
result:  the probability is $0.89$ for $c=1$ 
(independent of $\tau_0$ and $\tau$) and 
decreases as $c$ increases, tending to $0.5$ as $c\to\infty$. 

\begin{figure} 
\begin{center}
\includegraphics[height=4cm]{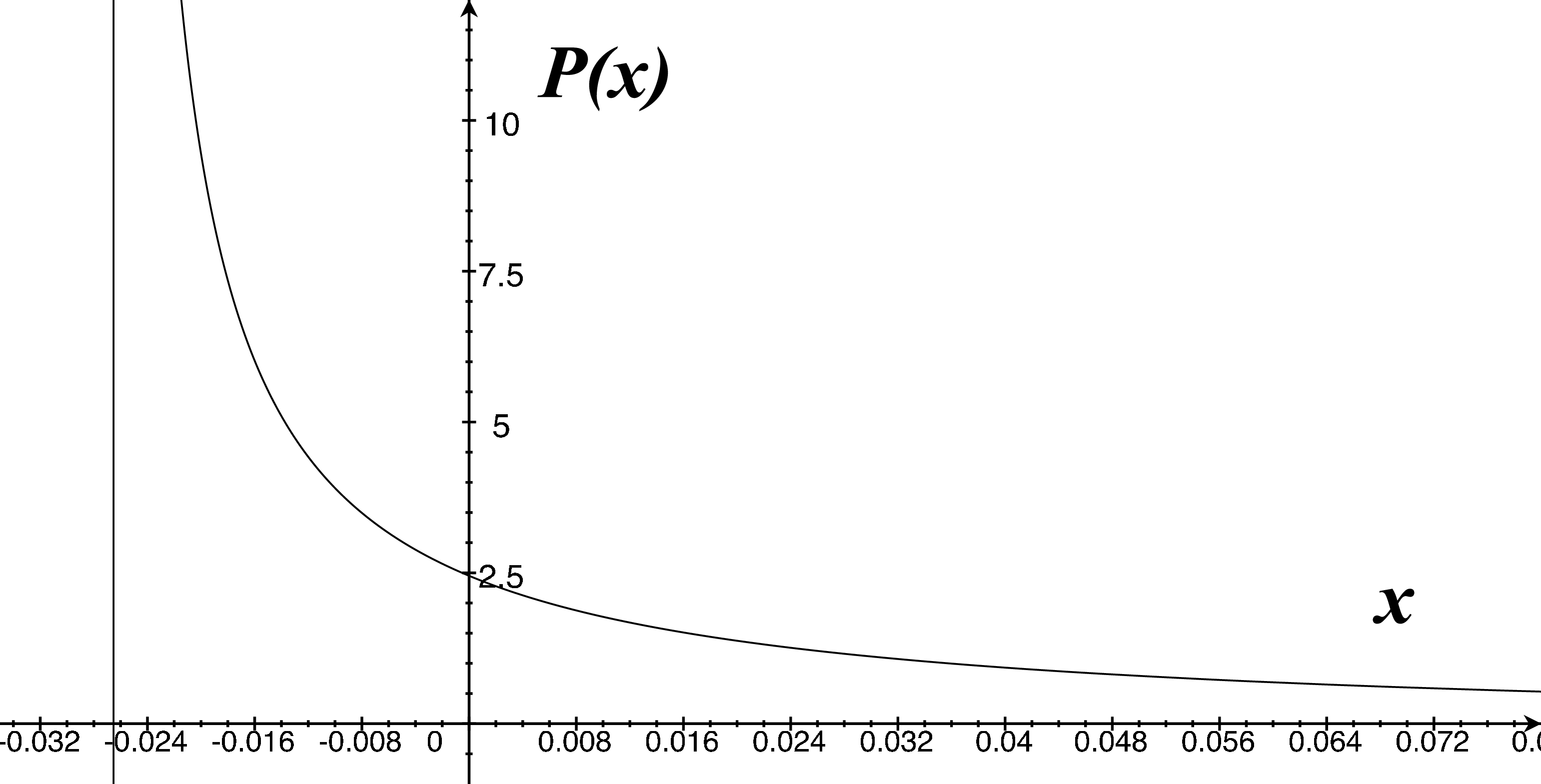}
\end{center}
\caption{The probability distribution $P(x)$ for the smeared energy density
  of a massless scalar field in two-dimensional Minkowski spacetime is
plotted. Here $x = \rho \, \tau^2$, where $\rho$ is the energy density
operator averaged in time with a Gaussian function of width
$\tau$. The lower limit of $P(x)$ occurs at $x=-x_0= -1/(12 \pi)$,
illustrated by the vertical line. }  
\label{fig:prob} 
\end{figure}

When both $T_R$ and $T_L$ are combined to obtain the overall probability
distribution for the energy density $T_{00}(t,0)$, the effect 
is to replace $c/24$ by $c/12$ in the probability distribution and 
definition of $\omega_0$. The averaged energy density operator is
$\rho =\frac1{\sqrt{\pi}\,\tau} 
\int_{-\infty}^{\infty} T_{tt}(t,0)\, {\rm e}^{-t^2/\tau^2} \, dt
\,.$
Then $x=\rho\tau^2$ is distributed according to the shifted Gamma distribution
with probability density function 
\begin{equation}
P(x) = \vartheta(x+x_0)\frac{\pi^{c/12}(x +x_0)^{c/12 -1}}{\Gamma(c/12)}\, 
{\rm e}^{-\pi(x+x_0)}\, ,  
\end{equation}
where $-x_0/\tau^2$ is the quantum inequality bound on
expectation values of $T_{tt}$ in arbitrary quantum states. The
probability distribution is plotted in Fig.~\ref{fig:prob} for the
case of a free massless scalar field in two-dimensional Minkowski
spacetime, for which $c=1$. In this case,
the probability of a negative result is $0.84$, and the corresponding 
QI bound coincides with that of \cite{Flanagan97}. 

This is a remarkable result which sheds light on the nature of vacuum
energy fluctuations, at least in two-dimensional spacetime. The zero
average value of the smeared energy density in the vacuum state
arises, in a sequence of measurements, from many negative values being
balanced by rarer, but larger positive values. 

Although the probability distribution for stress tensor operators in
four-dimensional spacetime is not yet known, we have been able to
construct distributions for the square of a massless scalar field
by empirical methods. This amounts to calculating several moments, and
then showing that they fit to a shifted Gamma distribution to high
accuracy. 
Using the Lorentzian sampling function $L(t)=\tau/(\pi(t^2+\tau^2))$,
we computed the first twenty moments of
$\int {:}\Phi^2{:}(t,\xb) L(t)\,dt$ in
the vacuum state, using symbolic integration in Maple. After suitable normalization, the computed moments are integers; the first eight of which are presented in 
Table~\ref{tab:Phi2L}. Remarkably, all twenty moments are fitted {\em exactly} 
by the shifted Gamma distribution Eq.~\eqref{eq:shifted_Gamma} 
with parameters
\begin{equation}
\alpha  =\frac{1}{72}, \qquad 
\beta    = \frac{4\pi^2\tau^2}{3}, \qquad
\omega_0 = \frac{1}{96\pi^2\tau^2}.
\end{equation}
Because the form in Eq.~(\ref{eq:shifted_Gamma}) is already normalized, we
need the $n= 1, 2,$ and $3$ moments to fit the above
parameters. However, as the remaining $17$ moments are reproduced
exactly, this provides strong evidence that this is the correct
distribution.

\begin{table}
\begin{center}
\begin{tabular}{|c||c|c|c|c|c|c|c|c|c|}\hline
$n$ & $0$ & $1$ & $2$ & $3$ & $4$ & $5$ & $6$ & $7$ & $8$ \\ \hline
$M_n$ &
$1$ & $0$ & $2$ &  $48$ & $1740$ &  $83904$ & $5051640$ &  $364724928$ & $30707616912$ \\
\hline
\end{tabular}
\end{center}
\caption{Computed moments $M_n = (4\pi\tau)^{2n}\GG_n[L]$ for 
the Lorentzian averaged squared field operator.}
\label{tab:Phi2L}
\end{table}

The lower bound  of the probability density support is 
$-\omega_0$, which provides a conjectured optimal QI bound
\begin{equation}
\int \ip{\psi}{{:}\Phi^2{:}(t,0)\psi} L(t)\,dt \ge -\frac{1}{96\pi^2\tau^2}
\end{equation}
for all normalized $\psi$ in the relevant domain. By comparison, 
the methods of \cite{FewsterEveson98} yield
a QI~\footnote{Compare with the QI 
given in Eq.~(5.5) of \cite{FewsterEveson98}; the 
starting point for Eq.~\eqref{eq:FEQI} is to 
set $p=1/\sqrt{\omega}$ in Eq.~(3.11) of \cite{FewsterEveson98}.} 
\begin{equation}\label{eq:FEQI}
\int 
\ip{\psi}{{:}\Phi^2{:}(t,\xb)\psi} f(t) dt \ge
-\frac{1}{8\pi^2}\int_{-\infty}^\infty 
\left[\frac{d}{dt}\sqrt{f(t)}\right]^2 dt.
\end{equation}
Substituting the Lorentzian $L$ 
in place of $f$, the lower bound is $-1/(64\pi^2\tau^2)$, 
which is weaker by a factor of $3/2$ than (and therefore does not
contradict) the conjectured bound just given.

Just as in the case of fluxes and energy densities in two dimensions,
the probability distribution diverges integrably at its lower limit.
In this case, the probability of a negative outcome for a measurement
of the averaged squared field is $0.95$.

The first fifteen moments of ${:}\Phi^2{:}$ smeared against the 
squared Lorentzian $S(t) = 2\tau^3/(\pi(t^2+\tau^2)^2)$ have also been computed
and are exactly fitted by the shifted Gamma distribution
\eqref{eq:shifted_Gamma} with parameters
$\alpha   ={1}/{45}$,
$\beta    = {8\pi^2\tau^2}/{15}$, and
$\omega_0 = {1}/({24\pi^2\tau^2})$.
The corresponding conjectured optimal QI bound is again stronger
by a factor of $3/2$ than that of Eq.~\eqref{eq:FEQI}; 
it is tempting to conjecture that the optimal bound
for general sampling functions is given by \eqref{eq:FEQI} with
a $12$ replacing the $8$. 

As in the case of the stress tensor in CFT with \hbox{$c<24$,} these
distributions for the squared scalar field are singular at the lower 
limit, although with an integrable singularity. The physical effect
of this singularity is to favor negative fluctuations close to the 
lower bound. Note that there nothing unphysical about this 
singularity, as the observable quantity, the probability of an outcome
in a finite interval, is finite.

We have also attempted a similar fit to the Lorentzian average
of the squared time derivative of a massless scalar field and to the
squared electric field. In both cases, a shifted Gamma distribution
which is fit to the lower moments underestimates the higher moments.
This suggests that the correct probability distribution for these
quantities, and for the four-dimensional stress tensor,
 has a positive tail which falls off more slowly than is
described by the shifted Gamma distribution. 

The probability distribution for the energy density in 
four-dimensional theories is of considerable
interest. One application is to inflationary cosmology, where
quantum stress tensor fluctuations might contribute a potentially
observable component to the cosmological density
fluctuations~\cite{WNF07}. This component would be non-Gaussian
in a way which is associated with the skewness of the quantum stress
tensor probability distribution. Another, perhaps more exotic,
application to cosmology arises in models which employ anthropic
reasoning to attempt to compute probabilities of various observables.
These models require a counting of observers,  possibly including
``Boltzmann brains'' which have nucleated from the vacuum in deSitter
or even Minkowski spacetime.   
If this is the more prevalent type of observer,  
it would greatly complicate attempts at anthropic
prediction. (For further discussion and references,
see, for example Refs.~\cite{GV08,DGLNSV}.)
The key to studying this question is in the details of the
long positive tail of the probability distribution.

\begin{acknowledgments}
This work was supported in part by the National
Science Foundation under Grants PHY-0855360 and PHY-0652904.
CJF thanks the Erwin Schr\"odinger Institute, Vienna, and
the organizers of the program {\em Quantum Field Theory on Curved Space-times
and Curved Target Spaces} for support
and hospitality during the final stages of this work.
\end{acknowledgments}


\end{document}